# AN EXPERIMENTAL APPARATUS FOR STUDY OF DIRECT β-RADIATION CONVERSION FOR ENERGY HARVESTING

Y. Haim[1], Y. Marciano[2] and G. deBotton[1]

[1]Department of Mechanical Engineering, Ben-Gurion University of the Negev, Beer-Sheva, Israel
[2]Department of Materials Engineering, Ben-Gurion University of the Negev, Beer-Sheva, Israel

**Abstract:** This paper introduces the development and testing of an experimental apparatus for characterization of a direct charging nuclear battery. The battery consists of a parallel-plates capacitor which is charged in a vacuum by the current of β-radiation particles (electrons) emitted from a radioisotope. A $^{63}$Ni radioisotope with an activity of 15mCi that produces a 20pA current was selected as the radiation source. The apparatus is unique in its design, having ultra-low leakage current and a few options for charge measurements. Preliminary results of a few tests are presented to demonstrate the capabilities of the apparatus.

## INTRODUCTION

In recent years there is an increased demand for long lasting miniature batteries. Among the many possible applications we mention a network of wireless sensors [1]. These sensors need to transmit data from one sensor to another and to a central data acquisition device. A major requirement for the development of such systems is a reliable long term and maintenance free portable power source. The lack of availability of such power source hinders the wide spread usage of these applications.

Radionuclide (radioactive material) is material with unstable nucleus which emits energy to become more stable. Radionuclides have long lives and extremely high energy density compared to other materials. In addition, their power density is not affected by temperature, pressure or other ambient conditions. Therefore, nuclear materials are good candidates for energy sources. However, energy conversion methods are needed to effectively and safely convert the nuclear energy into a useful electrical energy.

A few approaches for utilizing the energy stored in these materials have been devised in the past. Of special interest are the direct conversion procedures. Some radionuclides generate thermal energy, which is then converted into electricity. Accordingly, the energy conversion techniques are categorized as thermal like: thermionic and thermoelectric and as non-thermal like: beta-voltaic, photoelectric, and direct charge accumulation. However, the question which is the best conversion method for energy harvesting of radionuclides materials is still open.

This study is concerned with the direct charging nuclear battery. This method has been accomplished under one of three basic processes [2]: (1) primary current collector with vacuum insulation, (2) primary current collector with solid dielectric and (3) secondary current collector. In the first process, which is discussed in this study, a parallel-plates capacitor is charged by the current of β-particles. These particles are emitted from a radioactive layer deposited on one of the electrodes to the other electrode through a vacuum gap.

In 1913 Moseley [3] demonstrated the first direct conversion of radioactive decay energy into electricity by the use of a radium source. The source was mounted on an insulated rod at the center of a silver coated glass ball. An electric potential of 150kV was achieved before a voltage-breakdown. Moseley's generator was based on the emission of β-particles from an insulated radioactive source that becomes positively charged. In the early 1950's, in an effort to find an energy source for space applications, many studies on direct energy conversion of β-radiation were conducted. However, the efficiencies achieved were too low [2]. In the last decade the efficiencies of direct charging nuclear batteries increased significantly thanks to new energy harvesting concepts and technologies [4, 5]. However, further research is needed in order to narrow the gap between the theoretically predicted efficiency and those attained in practice.

This study is concerned with direct β-radiation conversion from a $^{63}$Ni source. A versatile dedicated experimental apparatus was designed and constructed for the measurements of charges, high electric potentials and currents at different gaps between the electrodes. We emphasize that the high electric potential is crucial for the utilization of the particles kinetic energy. The source produces a current in order of a few pA and hence we adhere to strict design concepts to minimize the leakage current.

## EXPERIMENTAL SETUP

The experimental apparatus is shown in Figs. 1 and 2. In Fig. 1 a schematic description of the apparatus is presented. The entire experiment is carried out inside a vacuum chamber. The vacuum chamber has a diameter of 300mm and a height of 450mm. The vacuum is established by a diffusion pump and a rotary pump to a base pressure of less than $1 \times 10^{-4}$ torr. The apparatus comprises of a variable parallel-plates capacitor formed by the metal plate upon which the radioactive layer is deposited and an aluminum collector plate whose electric potential is measured with an electrostatic non-contact voltmeter. The radioactive material installed on a Teflon mobile linear actuator. An electric motor is used as a switch to control the electric circuit.

The radionuclide source is a 15mCi $^{63}$Ni, with a cuboid shape whose dimensions are 38X10X0.25 mm. The source produces a 20pA current. The source is attached to a linear actuator operated by an Oriel DC encoder and controlled by a dual Oriel controller model 18009. The vacuum gap between the capacitor plates is determined according to the source position.

The electric circuit is controlled with a switch made out of a Teflon rod which is connected to an electric motor. A metallic wire with a metallic foil at its end is connected to the other side of the Teflon rod. This high insulation switch has two positions (Fig. 1). In the first position the foil is in contact with the aluminum collector and the capacitor is charged via an external power source or discharged to an electrometer for charge measurements. The second position is for the energy harvesting process, where the metal foil is rotated at a 180° angle relative to the first position and the collector is insulated from the environment. The reason for using the metallic foil is to ensure a good electric contact without application of mechanical loading on the collector.

In Fig. 2 an overview of the experimental chamber is presented. The collector plate is made of an aluminum disk, 50 mm in diameter, which is supported by three Teflon tubes. A Faraday shield is installed around the apparatus in order to reduce electrical noise. The voltage on the direct charged capacitor is determined by the non-contact USSVM (Ultra Stable Surface DC Voltmeter) sensor produced by Alpha Lab. The non-contact voltmeter sensor is located 25.4mm above the aluminum plate. A high voltage DC power supply, Model 215, BERTAN is used for calibration of the electric potential. The charges were measured by Keithley 610B electrometer. The instruments were connected to the components in the vacuum chamber through a hermetic electrical feed-through. The data was collected with a Fluke 2620A hydra data acquisition unit connected to a personal computer (PC).

The apparatus enables to charge the capacitor to a voltage of more than 3kV either by the radioisotope or by the external power source. The capacitor charge can be measured by discharging it to the electrometer. Two other possible measurements are of the charge leaving the source while the collector plate is grounded and of the accumulated charge on the collector plate while it is connected directly to the electrometer.

These features are attained with an extremely low leakage current thanks to a strict design involving Teflon tubes that support the collector plate.

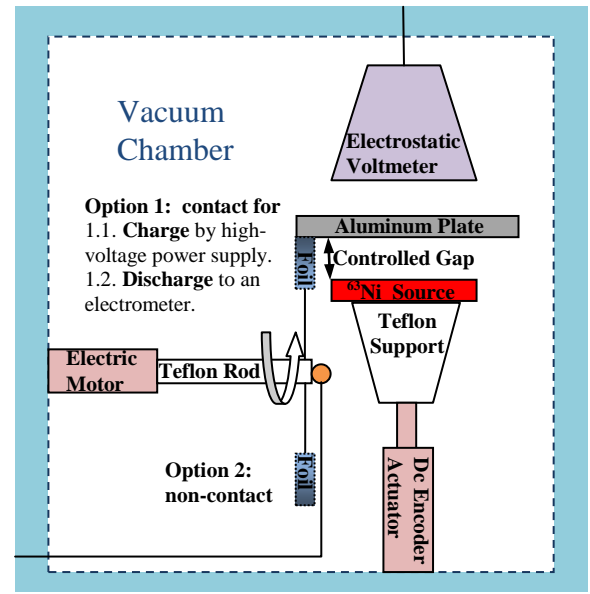

*Fig. 1: Schematic description of the measurement principle.*

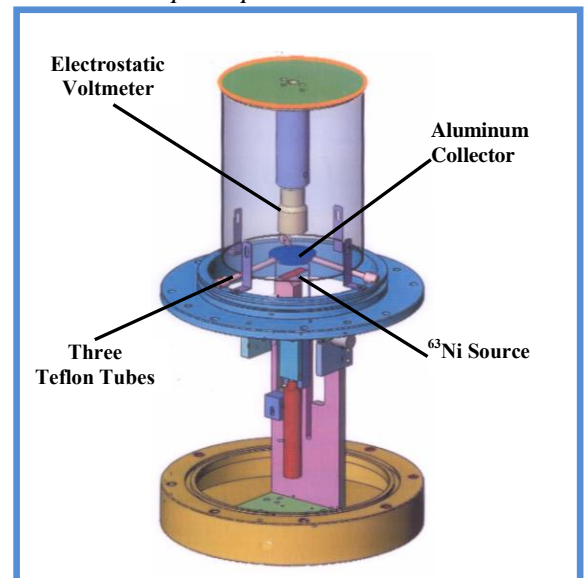

*Fig. 2: Overview of the experimental-chamber.*

## RESULTS AND DISCUSSION

The leakage current from the capacitor was tested by charging it, in the vacuum chamber, to an electric potential of 3kV and maintaining the collector plate insulated for 7 days. By discharging the capacitor to the electrometer at the end of this period it is found that the resistance of the electrical insulation is in the order of $10^{17}$ Ohms. This corresponds to a leakage current of 0.03pA at 3kV.

In order to determine the capacitance of the variable parallel-plates capacitor, it was charged to a few electric potentials at different gaps. Next, the charge was measured by discharging it to the electrometer. The electrometer readings are shown in Fig. 3 versus the input potential for the various gaps. In these measurements a metal cuboid plate whose dimensions are identical to those of the radioisotope source was used as the positive electrode.

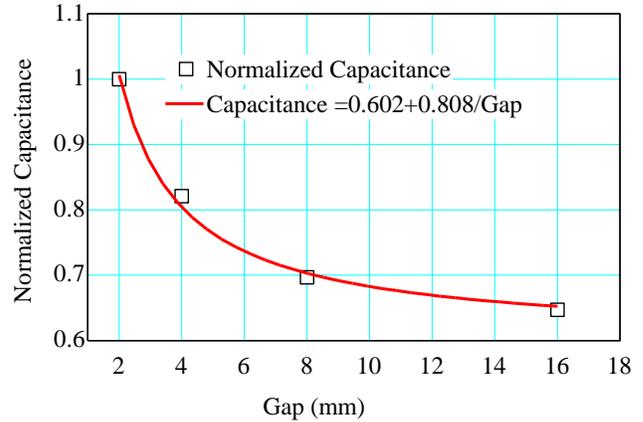

*Fig. 4: Normalized capacitance as a function of the gap.*

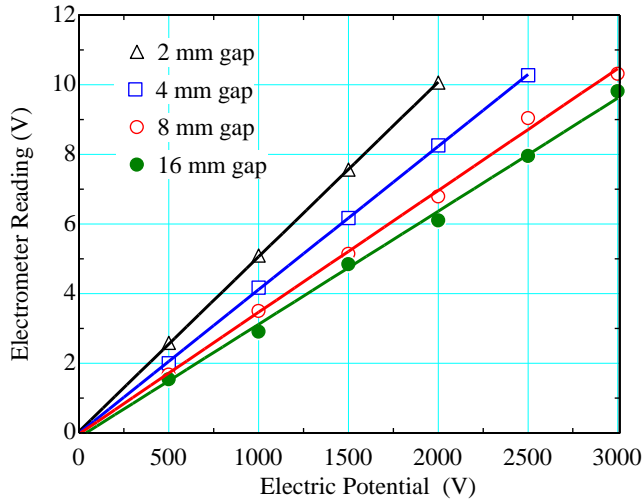

*Fig. 3: The measured charge at different gaps as a function of the electric potential.*

In Fig. 3 the linearity of the electric potential between the capacitor plates and the charge measured by the electrometer is shown. This linearity was verified for a few gaps between the capacitor plates. Based on these results the capacitance of the capacitor was calculated for each gap (Fig. 4).

In Fig. 4 the normalized capacitance is shown versus the gap between the electrodes. The experimental correlation for the capacitance is in agreement with corresponding analytical predictions.

Subsequent measurements were carried out with the $^{63}$Ni radiation source. The electric potential between the capacitor plates was measured versus time by the non-contact voltmeter for 2, 4, 8 and 16 mm gaps. Each test lasted 10 minutes. Curves of the electric potential versus the charge time are shown in Fig. 5.

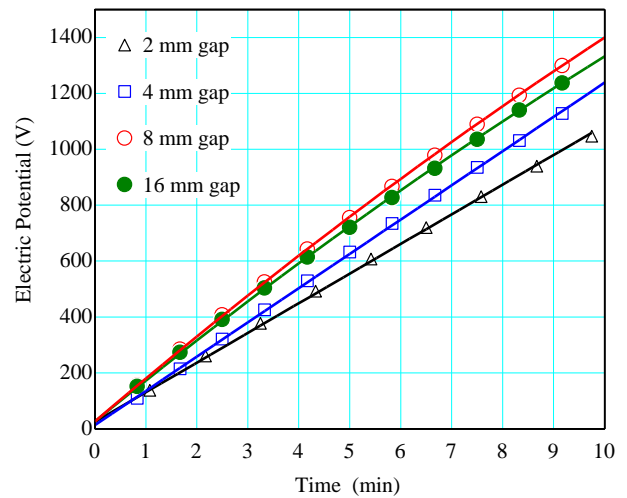

*Fig. 5: The attained electric potential of charging versus time for different gaps*

The maximal electric potential attained during these experiments was 1.4kV, at a 8 mm gap 10 min after the beginning of the test. The curves in Fig. 5 are almost linear. The rate at which the electric potential rises depends on the gap. Assuming a fixed flux of particles from the source, at small gaps the electric potential increases with the gap as expected. However, at a 16 mm gap the effect of the view factor between the source plate and the collector seems to be higher and we observe a decrease in the electric potential. The assessment of an optimal distance between the electrodes will be examined in future studies.

## CONCLUSION

An experimental apparatus for energy harvesting of a $^{63}$Ni source by direct charge accumulation was constructed and tested. The apparatus provides the following features: an ability to control the gap between the source and the collector, a non-contact measurement of the electric potential between the electrodes, a versatile measurement of the charge leaving the source and the charge accumulated on the collector, and the ability to calibrated the system by an external high voltage DC power supply. Tests with the external power supply demonstrated an extremely low leakage current that enables energy harvesting from a weak β-radiation source.

The experimentally determined correlation for the capacitance of the variable parallel-plates capacitor was verified against corresponding analytical predictions. In turn, the stored energy in the radioactive capacitor can be calculated from measurements of the electric potential between the capacitor plates.

The results of a few preliminary tests with a radioactive source are also shown. Up to a level of 1.4kV we find an almost linear increase of the electric potential with time. This suggests that backscattering of electrons from the collector plate is still negligible. The dependence of the slope of the curve on the gap is non-monotonous, implying that an optimal combination of gap and electric potential between the electrodes is to be identified. This optimization, among other tasks, will be studied with the aid of the devised apparatus.

**REFERENCES**
[1] Santi P. 2012 *Mobility models for next generation wireless networks: ad hoc, vehicular, and mesh networks*. (Hoboken, NJ: Wiley)
[2] Shorr W. 1955 Nuclear Batteries – A Survey *Peaceful Uses of Atomic Energy*, **15** 310-16
[3] Moseley H. G. J. 1913 The Attainment of High Potentials by the Use of Radium, *Proceedings of the Royal Society of London. Series A*, **88** 471-476
[4] Duggirala R., Lal A., Radhakrishnan S. 2010 *Radioisotope Thin-Film Powered Microsystems* (Springer)
[5] Kavetskiy A., Yakubova G., Yousaf S. M., Bower K. 2011 Efficiency of Pm-147 direct charge radioisotope battery, *Appl. Radiat. Isot*., **69** 744 -748